\documentclass[useAMS,usenatbib]{mn2e} 
\usepackage{graphicx} 
\usepackage{txfonts}

\title[GRB 070419A --- central engine tuning]{Evidence for energy injection and a fine-tuned central engine at optical wavelengths in GRB 070419A} 
\author[A. Melandri]{A. Melandri $^{1}$\thanks{E-mail: 
axm@astro.livjm.ac.uk}, C. Guidorzi $^{2,3,1}$,  S. Kobayashi$^{1}$, D. Bersier$^{1}$, C. G. Mundell$^{1}$, \newauthor 
P. Milne$^{4}$, A. Pozanenko$^{5}$, W. Li$^{6}$, A. V. Filippenko$^{6}$, Y. Urata$^{7}$, M. Ibrahimov$^{8}$, \newauthor 
I. A. Steele$^{1}$, A. Gomboc$^{9}$, R. J. Smith$^{1}$, N. R. Tanvir$^{10}$, E. Rol$^{10}$\\ 
$^{1}$Astrophysics Research Institute, Liverpool John Moores University, Twelve Quays House, Egerton Wharf, Birkenhead, CH41 1LD, UK.\\ 
$^{2}$Dipartimento di Fisica, Universit\`a di Ferrara, via Saragat 1, I-44100 Ferrara, Italy.\\ 
$^{3}$INAF-Osservatorio Astronomico di Brera, via Bianchi 46, I-23807 Merate (LC), Italy.\\ 
$^{4}$ Steward Observatory, University of Arizona, 933 North Cherry Avenue, Tucson, AZ 85721, USA.\\ 
$^{5}$ Space Research Insitute (IKI), 84/32 Profoyuznaya Str, Moscow 117997, Russia. \\ 
$^{6}$ Department of Astronomy, University of California, Berkeley, CA 94720-3411, USA.\\ 
$^{7}$ Department of Physics, Saitama University, Saitama 338-8570, Japan.\\ 
$^{8}$ Ulugh Beg Astronomical Institute, Tashkent 700052, Uzbekistan.\\ 
$^{9}$ Faculty of Mathematics and Physics, University of Ljubljana, Jadranska 19, SI-1000 Ljubljana, Slovenia.\\ 
$^{10}$ Department of Physics and Astronomy, University of Leicester, University Road, Leicester LE1 7RH, UK.}

\begin{document} 
\bibliographystyle{plainnat}

\date{} 
 
\pagerange{\pageref{firstpage}--\pageref{lastpage}} \pubyear{2006} 
 
\maketitle 
 
\label{firstpage} 
 
\begin{abstract} 
 
We present a comprehensive multiwavelength temporal and spectral 
analysis of the FRED GRB 070419A. The early-time emission in the 
$\gamma$-ray and X-ray bands can be explained by a central engine 
active for at least 250 s, while at late times the X-ray light curve 
displays a simple power-law decay. In contrast, the observed behaviour 
in the optical band is complex (from 10$^2$ up to 10$^6$ s).  We 
investigate the light curve behaviour in the context of the standard 
forward/reverse shock model; associating the peak in the optical light 
curve at $\sim$450~s with the fireball deceleration time results in a 
a Lorenz factor $\Gamma \approx 350$ at this time. In contrast, the 
shallow optical decay between 450 and 1500 s remains problematic, 
requiring a reverse shock component whose typical frequency is above 
the optical band at the optical peak time for it to be explained 
within the standard model. This predicts an increasing flux density 
for the forward shock component until t $\sim$ 4 $\times$ 10$^6$~s, 
inconsistent with the observed decay of the optical emission from t 
$\sim$ 10$^4$~s. A highly magnetized fireball is also ruled out due to 
unrealistic microphysic parameters and predicted light curve behaviour 
that is not observed. We conclude that a long-lived central engine 
with a finely tuned energy injection rate and a sudden cessation of 
the injection is required to create the observed light curves - 
consistent with the same conditions that are invoked to explain the 
plateau phase of canonical X-ray light curves of GRBs. 
 
\end{abstract} 
 
 
\section{Introduction} 
 
The temporal shape of the prompt emission of gamma-ray bursts (GRBs) 
can show a variety of profiles: from narrow and symmetric to wide and 
asymmetric pulses. In some cases a less energetic precursor is also 
detected, and in other cases a few overlapping pulses can take place 
for the entire duration of the $\gamma$-ray emission. It is true that 
many of those pulses (overlapping or single) detected for 
long-duration GRBs can be described by the superposition of ``fast 
rise exponential decay'' (FRED) profiles, one for each 
pulse. Moreover, several GRBs display only a single-shot FRED-like 
emission over the background in the $\gamma$-ray passband, that can be 
easily described by a simple Norris exponential model (Norris et 
al. 1996). In the context of the standard fireball model (Rees $\&$ 
M{\'e}sz{\'a}ros 1992) one may expect such GRBs to exhibit comparably 
simple behaviour in their afterglows at other wavelengths, and 
therefore be ideal testbeds for the model. For full temporal and 
spectral coverage, the predicted properties of the multiwavelength 
light curves have well-predicted shapes, depending on the relative 
contribution of the different components. 
 
In the X-ray band the temporal decay of many GRBs observed by {\it 
Swift} is well described by a canonical ``steep-shallow-steep'' decay 
(Tagliaferri et al. 2005; Nousek et al. 2006; Zhang et al. 2006), with 
superposed flares observed in $\sim$50\% of bursts (O'Brien et 
al. 2006; Chincarini et al. 2007; Falcone et al. 2007). The initial 
steep decay is interpreted as the result of the high-latitude emission 
or as the contribution of the reverse-shock emission (e.g., Panaitescu 
\& Kumar; Zhang et al. 2006), the shallow phase is consistent with 
long-lasting central energy activity (Zhang et al. 2006), and the late 
steep decay is evidence of decaying forward-shock emission, (i.e., the 
standard X-ray afterglow phase). 
 
In the optical band, observed light curves are expected to show a 
variety of shapes depending on the relative contribution of the 
forward- and reverse-shock emission (Kobayashi $\&$ Zhang 2003; Zhang, 
Kobayashi $\&$ M{\'e}sz{\'a}ros 2003) and the starting time of the 
observations. In particular, if the optical observations start when 
the reverse-shock contribution still dominates or when the central 
engine is still active, the detected temporal decay deviates from a 
simple power law (see fig. 1 in Melandri et al. 2008). Melandri et 
al. (2008) investigated the behaviour of the early decay phase in the 
optical and X-ray bands for 24 GRBs and classified them into four 
self-consistent groups based on the relative shapes observed in the 
two bands.  Although 14 of the GRBs were well described by the 
standard model, the remaining 10 required adaptations such as ambient 
density gradients or energy injections from long-lived central 
engines. In some cases, even these modifications were unable to fully 
explain the light-curve properties. 
 
GRB 070419A was particularly problematic, despite a simple FRED 
$\gamma$-ray profile. In this paper we present a comprehensive 
multiwavelength temporal and spectral study of GRB 070419A, including 
published and unpublished data from infrared to $\gamma$-ray bands, 
and use this extensive dataset to challenge the standard model. 
 
Throughout we use the following conventions: the power-law flux 
density is given as $F(\nu,t) \propto t^{-\alpha} \nu^{-\beta}$, where 
$\alpha$ is the temporal decay index and $\beta$ is the spectral 
slope; we assume a standard cosmology with $H_0 = 
70$~km~s$^{-1}$~Mpc$^{-1}$, $\Omega_{m} = 0.3$, and $\Omega_{\Lambda} 
= 0.7$; and all uncertainties are quoted at the $1\sigma$ confidence 
level (cl), unless stated otherwise.

\section{Observations} 
 
\subsection{{\it Swift}/BAT data} 
 
On 2007 April 19 at 09:59:26 UT the Burst Alert Telescope (BAT; 
Barthelmy et al. 2005) triggered on GRB 070419A (Stamatikos et 
al. 2007a), a dim long GRB with a duration of $T_{90} \approx 
110$~s. The $\gamma$-ray emission observed by BAT is a single-shot 
FRED light curve, with a total duration of a few hundred seconds. This 
event displayed an average $\gamma$-ray fluence ($\sim$ 5 $\times$ 
10$^{-7}$ erg cm$^{-2}$) and a peak photon flux lying at the low end 
of the distribution of {\it Swift} GRBs (Sakamoto et al. 2008). The 
redshift of the burst ($z=0.97$, Cenko et al. 2007) resulted in an 
isotropic energy estimate of $\sim 1.6 \times 10^{51}$~ergs in the 
15--150 keV observed bandpass (Stamatikos et al. 2007c).

\subsection{{\it Swift}/XRT data} 
 
Follow-up observations of the BAT error circle were performed with the 
X-ray Telescope (XRT; Burrows et al. 2005) starting about 113~s after 
the BAT trigger. A bright, uncatalogued, fading source was detected at 
$\alpha$(J2000) = 12$^h$10$^m$58.80$^s$, $\delta$(J2000) = 
+39$^\circ$55$'$32.4$''$ (Perri et al. 2007) with an uncertainty of 
2.2$''$. The light curve showed a rapid decay at early times followed 
(after 10$^{3}$~s) by a power-law decline with $\alpha = 1.2 \pm 
0.2$. The X-ray spectrum is well fitted by an absorbed power law with 
a photon index $\Gamma_X = 2.46 \pm 0.09$ and $N_{\rm H} = (1.9 \pm 
0.2) \times 10^{21}$ cm$^{-2}$ (Stamatikos et al. 2007c).

\subsection{Optical data} 
 
The afterglow of GRB 070419A was followed in the optical band from 
about 3 min and continued up to $\sim$18 hr after the burst 
event. Late-time observations were acquired with the 8.4-m Large 
Binocular Telescope (LBT) in the SDSS-$r$ filter and showed a bump in 
the light curve (Dai et al. 2007), compatible with a supernova bump 
(see Section 4.3 for a more detailed discussion of the possible 
supernova contribution). A log of the observations is given in Table 
\ref{obslog0}, where we report the starting and ending time of the 
observations with each facility.

 
\setcounter{table}{0} 
\begin{table} 
\begin{minipage}{226mm} 
 \caption{Log of the optical observations.}  
\label{obslog0} \begin{tabular}{@{}ccccc} 
\hline Telescope & Filters & $\Delta t_{\rm start}$ & $\Delta t_{\rm end}$ & Reference \\ 
 & & (min) & (min) & \\ 
\hline 
\hline 
Super-LOTIS & $R$ & 3.43 & 35.46 & This work\\ 
KAIT & $VRI$ & 5.24 & 57.66 & This work\\ 
P60 & $Ri'$ & 5.87 & 92.35 & Cenko et al. (2008)\\ 
Kuiper & $VR$ & 27.68 & 97.68 &This work\\ 
FTN & $BRi'$ & 39.43 & 101.46 & This work\\ 
Kiso & $BR$ & 51.05 & 112.24 &This work\\ 
Lulin & $R$ & 112.75 & 236.32 &This work\\ 
Maidanak & $R$ & 352.80 & 352.80 & This work\\ 
KPNO 4-m & $R$ & 1036.97 & 1036.97 &This work\\ 
LBT & $r'$ & 5328.00 & 44352.00 & Dai et al. (2007)\\ 
\hline 
\hline 
 \end{tabular} 
\end{minipage}{226mm} 
$\Delta t_{\rm start}$ and $\Delta t_{\rm end}$ refer to the first and 
last photometric measurement acquired with the corresponding 
telescope. A complete list of all the photometric points presented on 
this work is reported in Table \ref{obslog}. 
\end{table}

We collected, cross-calibrated, and analysed all the available optical 
data acquired by ground-based telescopes for this event. We calibrated 
the optical data using a common set of selected catalogued stars 
present in the field of view. SDSS pre-burst observation (Cool et 
al. 2007) has been used for $r'$ and $i'$ filters, USNO-B1 $R2$ and 
$B2$ magnitudes for $R$ and $B$ filters respectively, and Nomad $V$ 
magnitudes for the $V$ filter. LBT $r'$ magnitudes are reported in the 
$R$ band applying an average color term of $<R-r'> \approx -0.31$ mag, 
estimated from several field stars. In the same way, KAIT $I$-band 
magnitudes are given in the $i'$ band assuming an average color term 
of $<I-i'> \approx -0.93$ mag. 
 
Next, the calibrated magnitudes were corrected for the Galactic 
absorption along the line of sight ($E_{B-V} = 0.028$ mag; Schlegel et 
al. 1998); the estimated extinctions in the different filters are 
$A_B$ = 0.12 mag, $A_V$ = 0.09 mag, $A_R$ = 0.07 mag, and $A_i'$ = 
0.05 mag. Corrected magnitudes were then converted into flux 
densities, $F_{\nu}$ (mJy), following Fukugita et al. (1996). Results 
are summarized in Table \ref{obslog}.

 
\setcounter{table}{1} 
\begin{table*} \begin{center} 
\begin{minipage}{226mm} 
 \caption{Optical and infrared calibrated magnitudes for the afterglow 
 of GRB 070419A.} \label{obslog} \scriptsize 
 \begin{tabular}{@{}cccccc|cccccc} 
\hline  
\hline 
Filter & Telescope & $\Delta t$ & Exp. Time & Magnitude & $F_{\nu}$ & Filter &  
Telescope & $\Delta t$ & Exp. Time & Magnitude & $F_{\nu}$ \\ 
 & & (min) & (min)& & (mJy) &  & & (min) & (min)& & (mJy) \\ 
\hline  
\hline 
$R$ & S-Lotis & 3.436  & 2.5 & 19.60 $\pm$ 0.21 & 0.0476 $\pm$ 0.0092 & $B$ & FTN & 47.97 & 1.67 & 20.86 $\pm$ 0.30 & 0.0205 $\pm$ 0.0057\\        
$R$ & S-Lotis & 3.925  & 2.0 & 19.75 $\pm$ 0.23 & 0.0415 $\pm$ 0.0088 & $B$ & FTN & 56.25 & 2.0 & 21.62 $\pm$ 0.35  & 0.0102 $\pm$ 0.0033\\        
$R$ & S-Lotis & 7.597  & 1.0 & 18.45 $\pm$ 0.12 & 0.1375 $\pm$ 0.0152 & $B$ & FTN & 65.13 & 3.0 & 21.69 $\pm$ 0.29  & 0.0095 $\pm$ 0.0025\\        
$R$ & S-Lotis & 8.718  & 1.0 & 18.65 $\pm$ 0.12 & 0.1143 $\pm$ 0.0126 & $B$ & FTN & 80.09 & 5.0 & 21.90 $\pm$ 0.23  & 0.0078 $\pm$ 0.0016\\        
$R$ & S-Lotis & 10.889 & 2.0 & 18.76 $\pm$ 0.11 & 0.1033 $\pm$ 0.0104 & $B$ & FTN & 96.90 & 4.0 & 22.17 $\pm$ 0.30  & 0.0061 $\pm$ 0.0017\\        
$R$ & S-Lotis & 13.062 & 2.0 & 18.84 $\pm$ 0.11 & 0.0959 $\pm$ 0.0097 & $B$ & Kiso & 58.78 & 5.0 & 21.80 $\pm$ 0.28 & 0.0086 $\pm$ 0.0022\\        
$R$ & S-Lotis & 15.968 & 4.0 & 18.90 $\pm$ 0.08 & 0.0908 $\pm$ 0.0067 & $B$ & Kiso & 74.18 & 5.0 & 21.50 $\pm$ 0.24 & 0.0114 $\pm$ 0.0025\\        
$R$ & S-Lotis & 20.428 & 4.0 & 19.02 $\pm$ 0.08 & 0.0813 $\pm$ 0.0060 & $B$ & Kiso & 89.41 & 5.0 & 22.20 $\pm$ 0.48 & 0.0059 $\pm$ 0.0027\\        
$R$ & S-Lotis & 24.886 & 4.0 & 19.14 $\pm$ 0.14 & 0.0728 $\pm$ 0.0094 & $B$ & Kiso & 104.60 & 5.0 & 22.49 $\pm$ 0.48 & 0.0045 $\pm$ 0.0021\\       
$R$ & S-Lotis & 29.677 & 5.0 & 19.27 $\pm$ 0.10 & 0.0646 $\pm$ 0.0059 & & & & & & \\ 
$R$ & S-Lotis & 35.468 & 5.0 & 19.58 $\pm$ 0.13 & 0.0485 $\pm$ 0.0058 & $V$ & KAIT & 5.245 & 2.0 & 19.811 $\pm$ 0.55 & 0.0470 $\pm$ 0.0248\\       
$R$ & KAIT & 3.665 & 0.333 & 20.45 $\pm$ 0.88 & 0.0218 $\pm$ 0.0196 & $V$ & Kuiper & 36.553 & 1.0 & 19.961 $\pm$ 0.11 & 0.0410 $\pm$ 0.0041\\    
$R$ & KAIT & 6.208 & 0.75 & 19.19 $\pm$ 0.17 & 0.0695 $\pm$ 0.0109 & $V$ & Kuiper & 38.388 & 1.0 & 19.857 $\pm$ 0.11 & 0.0451 $\pm$ 0.0045\\    
$R$ & KAIT & 9.733 & 1.0 & 18.59 $\pm$ 0.08 & 0.1208 $\pm$ 0.0089	& $V$ & Kuiper & 40.122 & 1.0 & 20.112 $\pm$ 0.12 & 0.0356 $\pm$ 0.0039\\    
$R$ & KAIT & 21.65 & 1.0 & 19.07 $\pm$ 0.14 & 0.0776 $\pm$ 0.0100	& $V$ & Kuiper & 41.858 & 1.0 & 20.144 $\pm$ 0.12 & 0.0346 $\pm$ 0.0038\\    
$R$ & KAIT & 24.00 & 1.0 & 19.29 $\pm$ 0.18 & 0.0634 $\pm$ 0.0105	& $V$ & Kuiper & 43.597 & 1.0 & 20.237 $\pm$ 0.13 & 0.0318 $\pm$ 0.0038\\    
$R$ & KAIT & 26.40 & 1.0 & 19.26 $\pm$ 0.18 & 0.0652 $\pm$ 0.0108	& $V$ & Kuiper & 45.336 & 1.0 & 20.141 $\pm$ 0.12 & 0.0347 $\pm$ 0.0038\\    
$R$ & KAIT & 30.572 & 3.0 & 19.57 $\pm$ 0.22 & 0.0490 $\pm$ 0.0099 & $V$ & Kuiper & 47.075 & 1.0 & 19.990 $\pm$ 0.11 & 0.0399 $\pm$ 0.0040\\    
$R$ & KAIT & 36.483 & 2.0 & 19.97 $\pm$ 0.25 & 0.0339 $\pm$ 0.0078 & $V$ & Kuiper & 48.826 & 1.0 & 20.009 $\pm$ 0.11 & 0.0392 $\pm$ 0.0039\\    
$R$ & KAIT & 41.173 & 5.0 & 20.15 $\pm$ 0.21 & 0.0287 $\pm$ 0.0055 & $V$ & Kuiper & 50.566 & 1.0 & 20.182 $\pm$ 0.13 & 0.0334 $\pm$ 0.0040\\    
$R$ & KAIT & 47.653 & 6.0 & 20.57 $\pm$ 0.30 & 0.0195 $\pm$ 0.0054 & $V$ & Kuiper & 52.315 & 1.0 & 19.956 $\pm$ 0.11 & 0.0412 $\pm$ 0.0041\\    
$R$ & KAIT & 57.661 & 7.0 & 20.71 $\pm$ 0.33 & 0.0171 $\pm$ 0.0053 & $V$ & Kuiper & 54.054 & 1.0 & 20.502 $\pm$ 0.17 & 0.0249 $\pm$ 0.0039\\    
$R$ & Kuiper & 27.681 & 1.0 & 19.54 $\pm$ 0.07 & 0.0503 $\pm$ 0.0032 & $V$ & Kuiper & 55.790 & 1.0 & 20.433 $\pm$ 0.16 & 0.0265 $\pm$ 0.0039\\    
$R$ & Kuiper & 29.420 & 1.0 & 19.50 $\pm$ 0.07 & 0.0522 $\pm$ 0.0033 & $V$ & Kuiper & 57.524 & 1.0 & 20.774 $\pm$ 0.21 & 0.0194 $\pm$ 0.0037\\    
$R$ & Kuiper & 31.148 & 1.0 & 19.65 $\pm$ 0.08 & 0.0455 $\pm$ 0.0033 & $V$ & Kuiper & 59.257 & 1.0 & 20.387 $\pm$ 0.15 & 0.0277 $\pm$ 0.0038\\    
$R$ & Kuiper & 32.890 & 1.0 & 19.60 $\pm$ 0.07 & 0.0477 $\pm$ 0.0030 & $V$ & Kuiper & 97.687 & 12.0 & 21.663 $\pm$ 0.30 & 0.0085 $\pm$ 0.0024\\   
$R$ & Kuiper & 64.489 & 4.0 & 20.81 $\pm$ 0.10 & 0.0156 $\pm$ 0.0014 & & & & & & \\ 
$R$ & Kuiper & 69.23 & 4.0 & 20.83 $\pm$ 0.10 & 0.0153 $\pm$ 0.0014 & $I$ & KAIT & 7.695 & 1.0 & 19.56 $\pm$ 0.45 & 0.0571 $\pm$ 0.0243\\        
$R$ & Kuiper & 73.96 & 4.0 & 20.95 $\pm$ 0.10 & 0.0137 $\pm$ 0.0012 & $I$ & KAIT & 8.500 & 1.0 & 18.84 $\pm$ 0.25 & 0.1109 $\pm$ 0.0257\\        
$R$ & Kuiper & 78.71 & 4.0 & 21.03 $\pm$ 0.11 & 0.0128 $\pm$ 0.0012 &  $I$ & KAIT & 11.05 & 1.0 & 18.50 $\pm$ 0.18 & 0.1517 $\pm$ 0.0252\\        
$R$ & Kuiper & 83.46 & 4.0 & 21.26 $\pm$ 0.18 & 0.0103 $\pm$ 0.0017 & $I$ & KAIT & 13.40 & 1.0 & 18.59 $\pm$ 0.22 & 0.1397 $\pm$ 0.0285\\        
$R$ & Kuiper & 88.751 & 4.0  & 21.13 $\pm$ 0.12 & 0.0116 $\pm$ 0.0013 & $I$ & KAIT & 16.44 & 2.0 & 18.75 $\pm$ 0.11 & 0.1205 $\pm$ 0.0122\\        
$R$ & FTN & 39.430 & 1.0 & 20.20 $\pm$ 0.24 & 0.0274 $\pm$ 0.0061 & $I$ & KAIT & 22.32 & 3.0 & 18.99 $\pm$ 0.12 & 0.0966 $\pm$ 0.0107\\        
$R$ & FTN & 47.384 & 0.5 & 20.37 $\pm$ 0.16 & 0.0234 $\pm$ 0.0034	& $I$ & KAIT & 31.77 & 5.0 & 20.18 $\pm$ 0.48 &  0.0323 $\pm$ 0.0147\\       
$R$ & FTN & 51.914 & 1.0 & 20.40 $\pm$ 0.16 & 0.0228 $\pm$ 0.0033	& $i'$ & FTN & 46.644 & 0.667 & 20.73 $\pm$ 0.32 & 0.0194 $\pm$ 0.0058\\     
$R$ & FTN & 58.809 & 2.0 & 20.46 $\pm$ 0.12 & 0.0216 $\pm$ 0.0023	& $i'$ & FTN & 53.698 & 1.0 & 20.58 $\pm$ 0.17 & 0.0223 $\pm$ 0.0035\\      
$R$ & FTN & 68.706 & 3.0 & 20.92 $\pm$ 0.15 & 0.0141 $\pm$ 0.0019	& $i'$ & FTN & 60.615 & 2.0 & 20.90 $\pm$ 0.16 & 0.0166 $\pm$ 0.0024\\      
$R$ & FTN & 78.623 & 2.0 & 21.10 $\pm$ 0.20 & 0.0119 $\pm$ 0.0022	& $i'$ & FTN & 72.505 & 3.0 & 20.96 $\pm$ 0.14 & 0.0157 $\pm$ 0.0020\\      
$R$ & FTN & 88.642 & 3.0 & 21.30 $\pm$ 0.21 & 0.0099 $\pm$ 0.0019	& $i'$ & FTN & 81.405 & 2.0 & 20.96 $\pm$ 0.16 & 0.0157 $\pm$ 0.0023\\      
$R$ & FTN & 101.46 & 4.0 & 21.15 $\pm$ 0.15 & 0.0114 $\pm$ 0.0015 & $i'$ & FTN & 92.399 & 3.0 & 21.58  $\pm$ 0.30 & 0.0089 $\pm$ 0.0024\\     
$R$ & Lulin & 112.75 & 10.0 & 21.26 $\pm$ 0.17 & 0.0103 $\pm$ 0.0016  & & & & & & \\ 
$R$ & Lulin & 125.31 & 15.0 & 21.70 $\pm$ 0.14 & 0.0068 $\pm$ 0.0009 & $J$ & UKIRT & 39.24 & 3.0 & 18.27 $\pm$ 0.08 & 0.0802 $\pm$ 0.0059\\ 
$R$ & Lulin & 236.32 & 30.0 & 22.43 $\pm$ 0.17 & 0.0035 $\pm$ 0.0005 & $J$ & UKIRT & 48.84 & 3.0 & 18.56 $\pm$ 0.09 & 0.0614 $\pm$ 0.0051\\ 
$R$ & Kiso & 51.05 & 5.0 & 20.13 $\pm$ 0.11 & 0.0292 $\pm$ 0.0029 & & & & & & \\ 
$R$ & Kiso & 66.65 & 5.0 & 20.75 $\pm$ 0.20 & 0.0165 $\pm$ 0.0030 & $H$ & UKIRT & 44.04 & 3.0 & 17.83 $\pm$ 0.13 & 0.0766 $\pm$ 0.0092\\ 
$R$ & Kiso & 81.88 & 5.0 & 20.90 $\pm$ 0.23 & 0.0144 $\pm$ 0.0030 & & & & & & \\ 
$R$ & Kiso & 112.24 & 15.0 & 21.66 $\pm$ 0.34 & 0.0071 $\pm$ 0.0022 & $K$ & UKIRT & 53.76 & 3.0 & 17.26 $\pm$ 0.18 & 0.0838 $\pm$ 0.0139\\ 
$R$ & Maidanak & 352.8 & 55.0 & 22.45 $\pm$ 0.32 & 0.0034 $\pm$ 0.0010 & $K$ & UKIRT & 99.90 & 15.0 & 18.30 $\pm$ 0.18 & 0.0321 $\pm$ 0.0053\\ 
$R$ & KPNO 4-m & 1036.97 & 5.0 & 23.37 $\pm$ 0.20 & 0.0014 $\pm$ 0.0003 & & & & & & \\  
\hline \hline 
 \end{tabular} 
\end{minipage} 
Magnitudes are not corrected for Galactic absorption. P60 (Cenko et 
al. 2008) and LBT (Dai et al. 2007) magnitudes are not reported in 
this table, but they are plotted in Fig. \ref{figopt4}. KAIT-$I$ 
magnitudes have been calibrated against SDSS-$i'$ catalogued 
magnitudes. Flux densities ($F_{\nu}$) have been estimated from 
extinction-corrected magnitudes. 
\end{center} 
\end{table*}

The optical afterglow of GRB 070419A was also detected by the {\it 
Swift}/UVOT (Roming et al. 2005). Observations began $\sim$115~s after 
the event. The best position for the optical afterglow is measured in 
the KAIT images at $\alpha$(J2000) = 12$^h$10$^m$58.82$^s$, 
$\delta$(J2000) = +39$^\circ$55$'$33.92$''$ (Chornock et 
al. 2007). Deriving accurate photometry of the afterglow was 
difficult, due to the presence of a diffraction spike from a mag 7 
star in the field of view. A clear detection was possible in the $V$ 
band, while only a 3$\sigma$ upper limits was estimated for the $B$ 
and $U$ filters (Stamatikos et al. 2007c). UVOT-$V$ magnitudes are 
plotted together with all of the ground-based optical data 
(Fig. \ref{figopt1} in Section 3.5), but no cross-calibration between 
UVOT and other optical data was performed. Thus we cannot exclude the 
presence of a large offset between UVOT magnitudes and calibrated 
ground-based photometry.

\section{Results} 
 
We have undertaken a complete temporal and spectral analysis of the 
available {\it Swift} data. In this section we report the results of 
our $\gamma$-ray/X-ray/optical analysis.

\subsection{Gamma rays} 
 
\subsubsection{Gamma-ray light curve} 
 
As observed by BAT, the $\gamma$-ray behaviour of GRB 070419A is a 
single FRED light curve, lasting a few hundreds seconds. It can be 
easily fit with a Norris simple exponential model (peakedness fixed to 
1): 
 
\begin{eqnarray} 
F(t) & = & N^{\rm BAT} \times e^{-|t-t_{\rm peak}|/t_{\rm rise}} ~~~~~~t < t_{\rm peak} \nonumber\\ 
     & = & N^{\rm BAT} \times e^{-(t-t_{\rm peak})/t_{\rm decay}} ~~~~t > t_{\rm peak}. 
\end{eqnarray} 
 
\begin{figure} \centering  
\includegraphics[height=6cm,width=8.5cm]{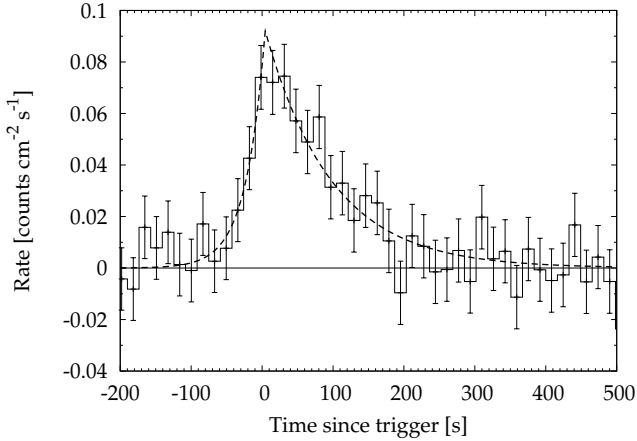} 
\caption{The BAT light curve. There is significant emission above  
the background up to $t \approx 300$~s. The dashed line represents the 
best-fitting of the light curve with the Norris profile.} 
\label{figbat}  
\end{figure}

%
%
\noindent The parameters of the best-fitting are $t_{\rm peak}^{\rm BAT}  
= 4.4 \pm 3.4$~s, $t_{\rm rise}^{\rm BAT} = 27.3 \pm 5.6$~s, and 
$t_{\rm decay}^{\rm BAT} = 93.0 \pm 11.0$~s ($\chi^{2}$/dof=0.49). The 
BAT light curve visible in Fig.~\ref{figbat} shows significant 
emission above the background up to 300 s after the burst onset 
time. In the same figure we show the result of the fit of the light 
curve done in the interval $-$100 to 300~s. 
 
 
\subsubsection{Gamma-ray spectral analysis} 
 
We independently analysed BAT data with the standard BAT pipeline 
(Krimm et al. 2004). Using the ftool ``battblocks'' (v1.7) we 
determined a value of $T_{90}$ = 112 $\pm$ 2~s (going from $-26$~s to 
+86~s). Fitting the 15--150 keV spectrum with a single power law, 
integrated over the $T_{90}$ interval, gives $\Gamma_{\rm \gamma}$ = 
2.4 $\pm$ 0.3, fluence $f$ = 5.1 $\pm$ 0.8 $\times$ 10$^{-7}$ erg 
cm$^{-2}$, and a corresponding peak photon flux of 0.14 $\pm$ 0.3 ph 
cm$^{-2}$ s$^{-1}$. All of these results are in agreement with the BAT 
team's published values (Stamatikos et al. 2007b,c). 
 
\subsection{X-rays} 
 
\subsubsection{X-ray light curve} 
 
\begin{figure} \centering  
\includegraphics[height=6cm,width=8.5cm]{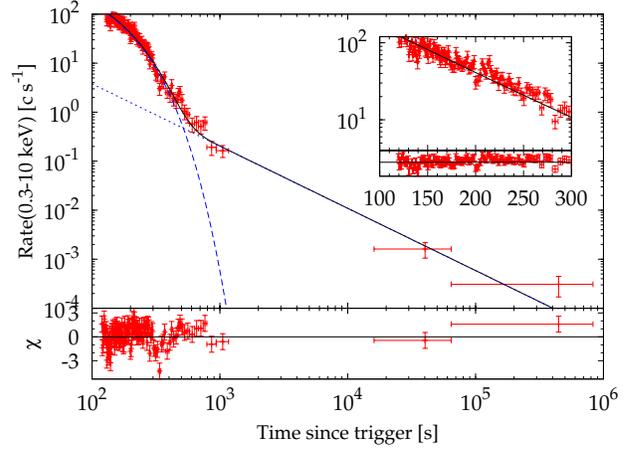} 
\caption{The XRT light curve, fitted using the exponential plus single  
power law model as explained in the text. The two components of the 
model are shown separately. In the inset the fit of the first 300~s of 
data with the exponential function is shown together with the 
residuals of the fit.} 
\label{figxrt2}  
\end{figure}

From our independent analysis of the XRT data we find that the light 
curve in the X-ray band can be fitted by an exponential plus power-law 
model, 
 
\begin{centering} 
\begin{equation} 
F(t) = N^{\rm XRT}_{\rm exp} \times e^{-(t-t_0)/\tau} + N^{\rm XRT}_{\rm pl}  
\times t^{-\alpha_X}, 
\end{equation} 
\end{centering} 
 
%
%
 
\noindent with best-fitting parameters (uncertainties are 90$\%$ cl)  
$t_0 = 0.0$~s (fixed, as insensitive parameter), $\tau = (71.9 \pm 
2.4)$~s, and $\alpha_X = 1.27^{\tiny -0.12}_{\tiny +0.18}$. The result 
of the fit can be seen in Fig. \ref{figxrt2}, where the two components 
of the fit are shown separately. These values are in good agreement 
with those found by Stamatikos et al. (2007c). However, the $\chi^{2}$ 
per degree of freedom (dof) of this fit is not acceptable 
($\chi^{2}$/dof = 190/127 $\approx 1.5$). 
 
 
A more complex model (exponential with peakedness free to vary, plus 
two power laws) gives a satisfactory result. The justification for a 
more complex model is the unacceptable $\chi^{2}$ of the previous fit, 
due to bad residuals of the early-time X-ray data. The best-fitting 
parameters are $N^{\rm XRT}_{\rm exp} = 32.1^{\tiny -7.5}_{\tiny 
+13.4}$ counts s$^{-1}$, $t_0 = 0.0$~s (fixed), $\tau = 259^{\tiny 
-18}_{\tiny +13}$~s, peakedness = $5.6^{\tiny -1.6}_{\tiny +2.2}$, 
$N^{\rm XRT}_{\rm pl1} = 14.96^{\tiny -5.58}_{\tiny +9.53} \times 
10^{7}$ counts s$^{-1}$, $N^{\rm XRT}_{\rm pl2} = 1.6 \pm 0.7$ counts 
s$^{-1}$, $\alpha_{\rm X,1} = 3.0 \pm 0.1$, and $\alpha_{X,2} = 
0.65^{\tiny -0.39}_{\tiny +0.35}$. With this more complex model the 
$\chi^{2}$ is now acceptable ($\chi^{2}$/dof = 137/124 $\approx$ 
1.1). Evidently, the improvement in the total $\chi^{2}$ is too large 
(53) compared with the change in the degrees of freedom (3), when 
moving from the first to the second model. The P-value associated with 
this change in the $\chi^{2}$ with 3 dof is $2 \times 10^{-11}$, so 
completely negligible, in agreement with what one would obtain with an 
F-test. However, it should be noted that the goodness of the fit in 
this case is related to the sparse data coverage at very late times. 
 
 
\begin{figure} \centering  
\includegraphics[height=5.5cm,width=8.3cm]{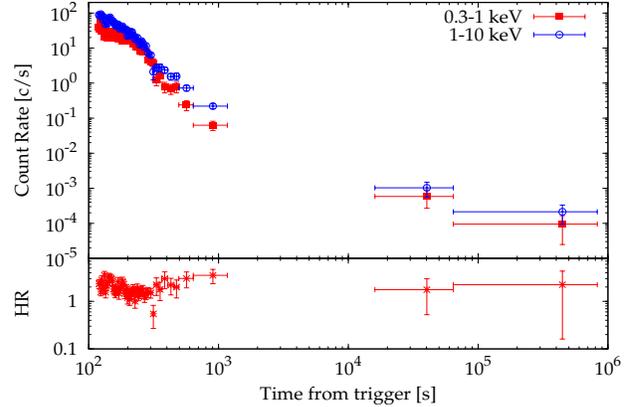} 
\caption{Hardness ratio (HR) and X-ray light curve in two different  
bands. At early times a spectral softening of the emission is visible 
over the first 200~s of data.} 
\label{figxrt4} 
\end{figure} 
 
In Fig. \ref{figxrt4} the X-ray light curve extracted in two separate 
bands (0.3--1 keV and 1--10 keV) is shown, together with the hardness 
ratio (hard/soft, lower panel) in the X-ray band. At early times 
(between $\sim$100 and $\sim$300~s) the X-ray emission softened, then 
hardened up to $\sim$1000~s when the hardness ratio became roughly 
constant up to the end of the observations.

\subsubsection{X-ray spectral analysis} 
 
 
\begin{figure*}
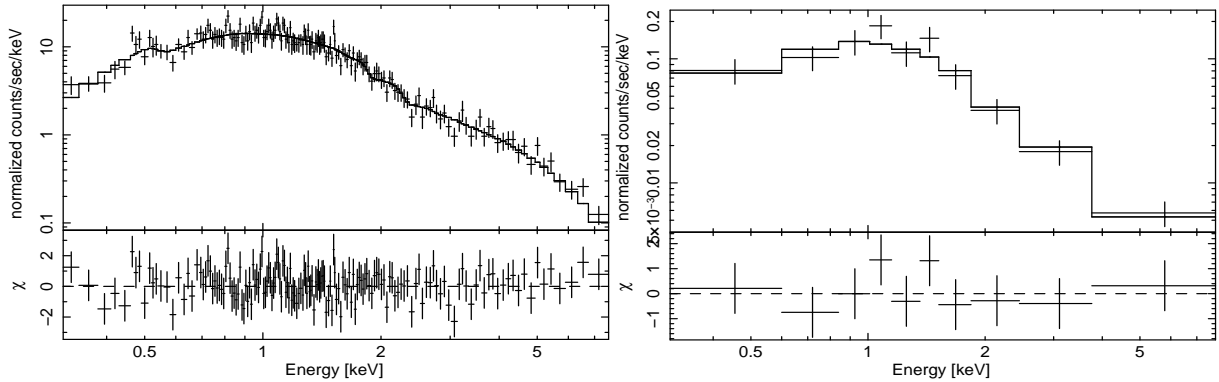
 \centering  
\includegraphics[height=8.0cm,width=5cm,angle=-90]{fig4a.eps} 
\includegraphics[height=8.0cm,width=5cm,angle=-90]{fig4b.eps} 
\caption{{\it Left panel}: XRT WT spectrum. {\it Right panel}:  
XRT PC spectrum.} 
\label{figxrt5} 
\end{figure*} 
 
In Fig. \ref{figxrt5} the total Windowed Timing (WT) 0.3--10 keV 
spectrum (left panel) and the total Photon Counting (PC) spectrum 
(right panel) are shown. The adopted model is an absorbed power 
law. The Galactic absorption is taken into account separately and the 
intrinsic $N_{\rm H}$ is given in the GRB rest frame. All 
uncertainties are at the 90$\%$ cl. 
 
The parameters of the fit for the WT spectrum (119--309~s) are $N_{\rm 
H(Gal)} = 2.4 \times 10^{20}$ cm$^{-2}$ (fixed), $N_{\rm H,z} = (5.1 
\pm 0.6) \times 10^{21}$ cm$^{-2}$, and $\Gamma_X = 2.2 \pm 0.1$ 
($\chi^{2}/dof = 142/161$). The parameters of the fit for the PC 
spectrum (310--1165~s) are $N_{\rm H(Gal)} = 2.4 \times 10^{20}$ 
cm$^{-2}$ (fixed), $N_{\rm H,z} = 3.4^{\tiny -2.5}_{\tiny +3.4} \times 
10^{21}$ cm$^{-2}$, and $\Gamma_X = 2.0 \pm 0.3$ ($\chi^{2}$/dof = 
4.8/7).

\subsubsection{X-ray temporally resolved analysis}  
 
\begin{figure} \centering  
\includegraphics[height=5.5cm,width=8.2cm]{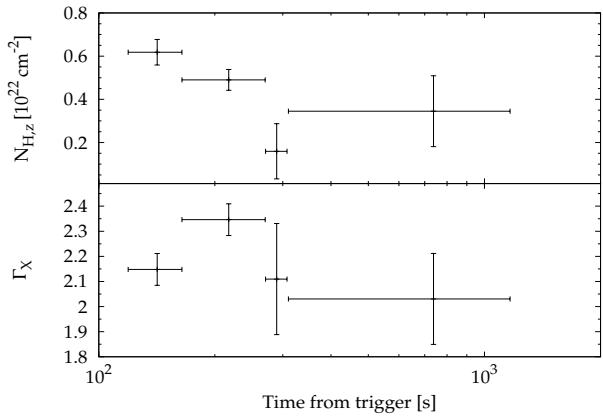} 
\caption{Intrinsic $N_{\rm H}$ and photon index as a function of  
time. Error bars shown are 1$\sigma$.} 
\label{figxrt7} 
\end{figure}

We extracted the XRT spectra in four separate time intervals each 
collecting ~2000 source photons. In Fig. \ref{figxrt7} we plot the 
$N_{\rm H}$ (intrinsic) and the photon index as a function of time. 
Clearly, as also shown in Fig. \ref{figxrt4}, there is marginal 
evidence for a hard-soft-hard trend in the photon index evolution; the 
same holds for the $N_{\rm H}$ evolution.

 
\subsection{Combined gamma/X-ray analysis} 
 
 
\begin{figure} \centering  
\includegraphics[height=5.5cm,width=8.2cm]{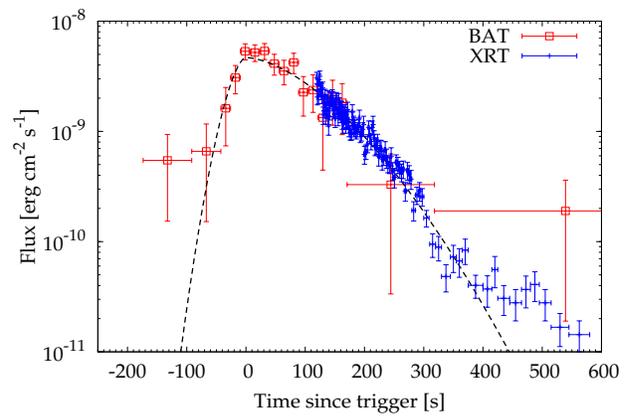} 
\caption{The joint BAT-XRT light curve can be fitted with the  
same Norris profile used for the BAT emission. See Section 3.3 for 
details.} 
\label{xrtbat1} 
\end{figure}

We calculated the flux in the 0.3--10 keV band after removing the 
$N_{\rm H}$ at low energies, both Galactic and intrinsic. Then we 
extrapolated it into the BAT 15--150 keV band taking into account the 
$\Gamma$ = 2.2 unbroken power-law spectrum between XRT and BAT. The 
result is shown in Fig. \ref{xrtbat1}, in which the initial 
exponential X-ray decay perfectly matches the decay of the FRED prompt 
emission seen by BAT. In this case we allowed the peakedness of the 
Norris profile free to vary. The best-fitting shown is obtained by 
fitting both fluxes together (90\% cl) with $t_{\rm peak}^{\rm 
BAT+XRT} = 0 \pm 15$~s, $t_{\rm rise}^{\rm BAT+XRT} = 36^{\tiny 
-24}_{\tiny +30}$~s, $t_{\rm decay}^{\rm BAT+XRT} = 147^{\tiny 
-17}_{\tiny +20}$~s, and peakedness = $1.64^{\tiny -0.16}_{\tiny 
+0.15}$ ($\chi^{2}$/dof = 163/128 $\approx$ 1.3). The fit was done on 
the points earlier than $t = 400$~s, after which the power law takes 
over (not shown in this plot). It is clear that the initial steep 
decay seen in the X-ray band is just the tail of the exponential 
(single shot) decay observed in the $\gamma$-ray band, corresponding 
to the tail of the prompt emission.

\subsection{The choice of $t_0$} 
 
 

The decay index of early afterglows is very sensitive to the choice of 
$t_0$. Correctly choosing $t_0$ is essential to derive the right index 
as well as to understand the emission process (Piro et al 2005; 
Tagliaferri et al. 2005; Quimby et al. 2006). In the previous section, 
$t_0$ is set at the GRB trigger time, and it is almost at the peak of 
the prompt emission. If the emission is due to an internal shock or 
external shock, $t_0$ should be set before the peak (Lazzati \& 
Begelman 2006; Kobayashi \& Zhang 2007). 
 
We can see how the choice of $t_0$ affects the decay index as follows 
(e.g., Yamazaki 2008). Let $t$ be the time since the GRB trigger; the 
peak is located at $t=0$~s on this time scale. Next, assuming that the 
temporal decay right after the peak is actually described by a power 
law with another time ($T=t+t_0$), where the interval between the time 
$T=0$ and $t=0$ is assumed to be exactly $t_0$, we get 
 
\begin{centering} 
\begin{equation} 
  f \propto T^{-\alpha} \propto (t+t_0)^{-\alpha}. 
\end{equation} 
\end{centering} 
 
\noindent Doing this, one finds that the flux $f$ is constant if $t\ll t_0$, 
while it is described by a power law $f \propto t^{-\alpha}$ if $t\gg 
t_0$. 
 
Although we have fit the early BAT-XRT data with an exponential 
function, it might be possible to fit the same data by a single power 
law with a different value of $t_0$. We tested this possibility by 
reexamining the data assuming different values for $t_0$. The decay 
indices right after the peak are $\alpha= 2.1 \pm 0.13$ if $t_0$ = 
100~s, $\alpha=3.2 \pm 0.2$ if $t_0$ = 200~s, and $\alpha=4.2 \pm 0.3$ 
if $t_0$ = 300~s. 
 
 
In all the cases, the light curves still have a clear break around the 
penultimate BAT point (corresponding to $T \approx t_0 + 250$~s, for 
any chosen value of $t_0$). With an even larger $t_0$, it is possible 
to describe the light curve roughly with a single power law. However, 
the best-fitting value of $\alpha$ is already very high with $t_0 = 
300$~s and it would be even higher for a larger $t_0$. The upper limit 
on the decay index is given by $\alpha=2+\beta$ (the high-latitude 
emission), where $\beta$ is the spectral index. The best-fitting value 
for $t_0 = 300$~s ($\alpha=4.2 \pm 0.3$) is already greater than this 
upper limit. Furthermore, after the break, the decay is even steeper 
($\alpha = 6.8 \pm 0.8$ or higher for correspondingly larger values of 
$t_0$). Thus, the post-break index is steeper than the limit from the 
high-latitude emission. 
 
The early BAT-XRT light curve is described neither by the emission 
from an external shock nor by that from a single internal shock. The 
very steep decay between $t=250$~s and $t=300$~s indicates that the 
central engine is active at least for $\sim$250~s, and that the early 
part ($t < 250$~s) should be the result of the superposition of many 
pulses (internal shocks).  Late-afterglow modeling is insensitive to 
the choice of $t_0$. In the rest of the paper (discussion on 
intermediate/late-time afterglow) we assume $t_0 = 0$~s.

\subsection{Optical Light Curve}

 
\begin{figure} \centering  
\includegraphics[height=5.2cm,width=6.7cm]{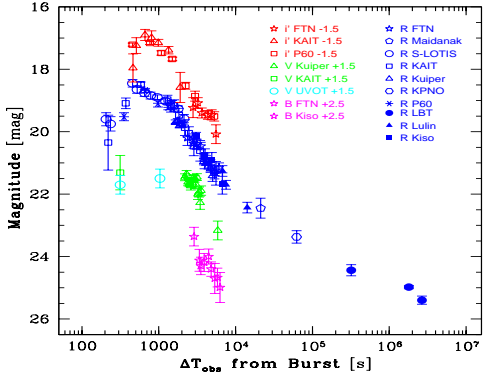} 
\caption{Observed optical light curves ($BVRi'$) for the afterglow of  
GRB 070419A.} 
\label{figopt1} 
\end{figure}

 
\begin{figure} \centering  
\includegraphics[height=5.5cm,width=8.2cm]{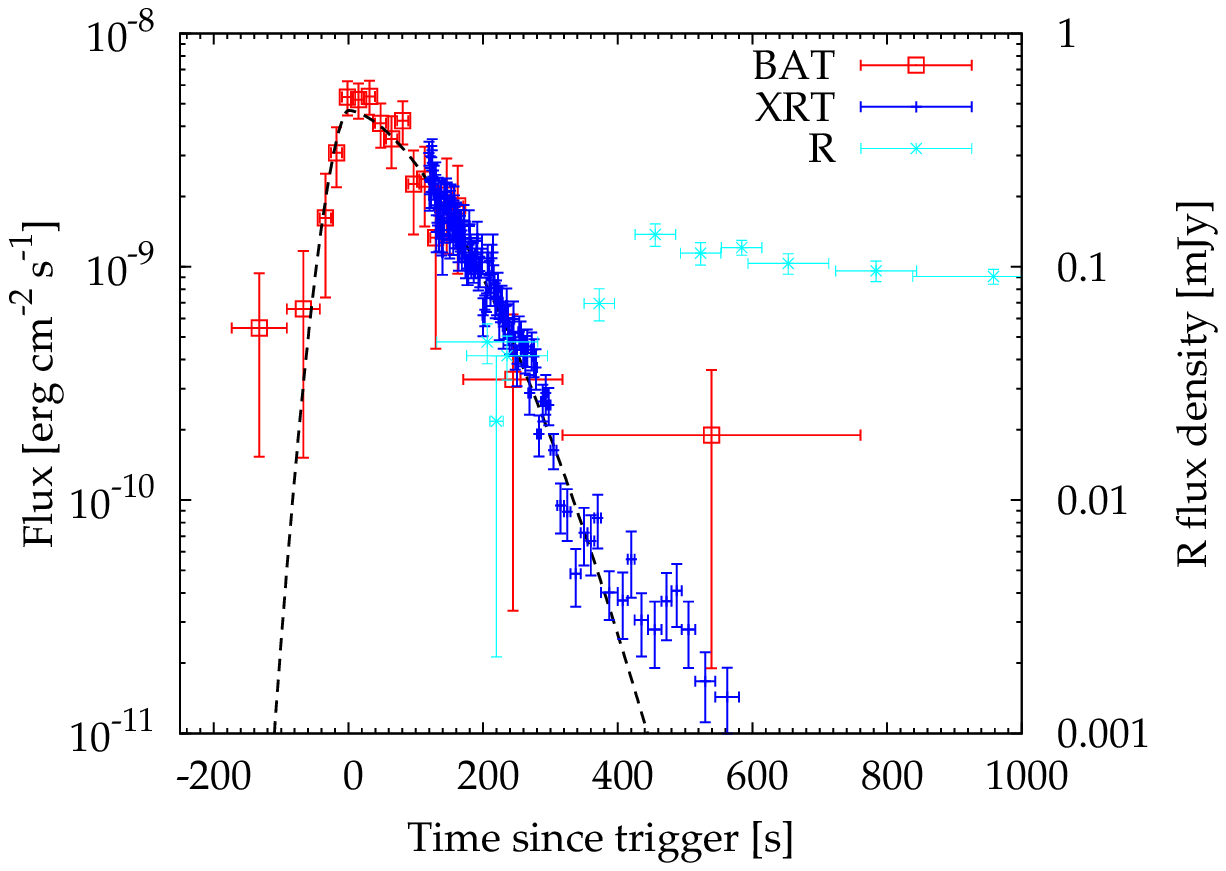} 
\includegraphics[height=5.5cm,width=8.2cm]{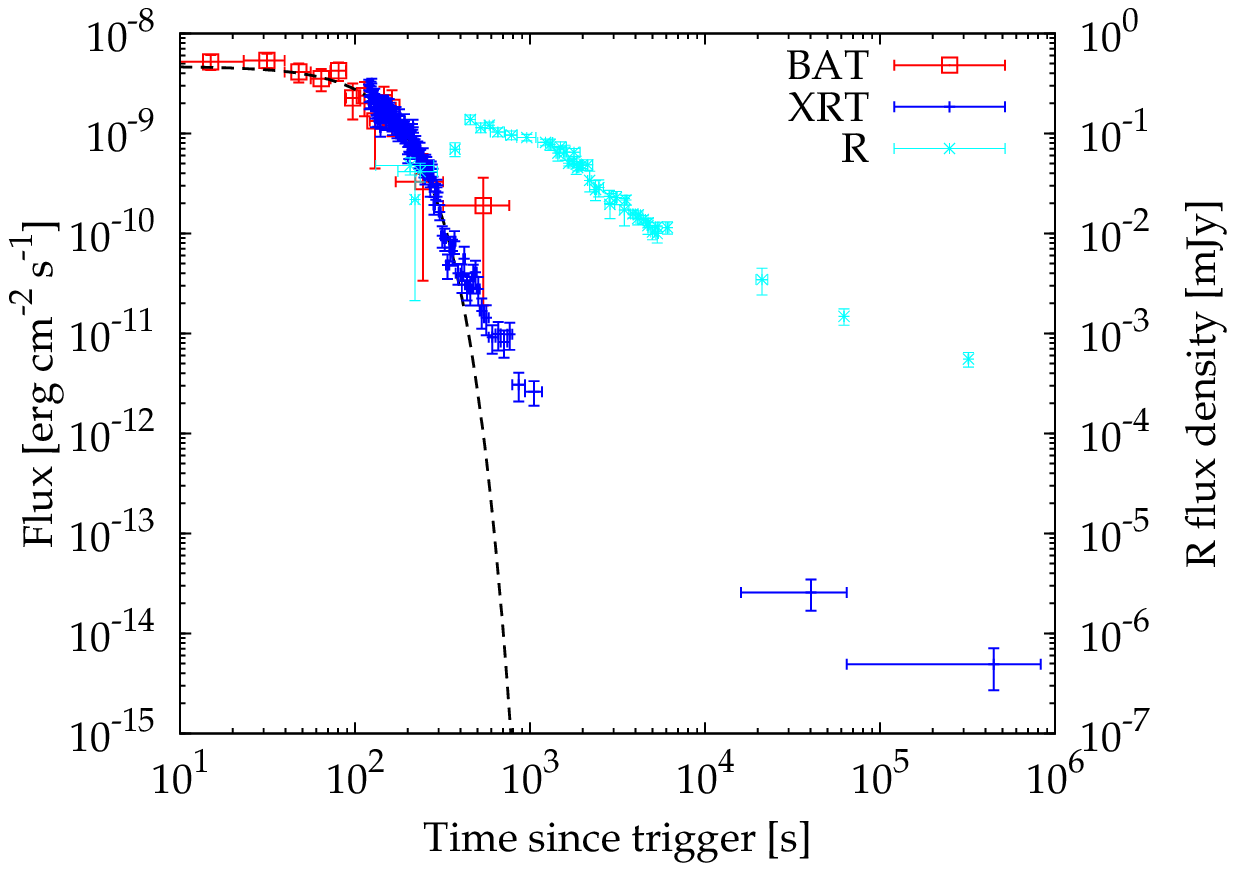} 
\caption{{\it Top panel}: $\gamma$-ray/X-ray/optical light curves on a  
linear-log scale. Same as in Fig. \ref{xrtbat1} for the joint BAT-XRT 
light curve. We overplot the optical flux in the $R$ band. {\it Bottom 
panel}: $\gamma$-ray/X-ray/optical light curves on a log-log scale. In 
this case we excluded from the plot the initial fluxes in the 
$\gamma$-ray (at negative times) and show the late-time optical and 
X-ray behaviours.} 
\label{figopt2} 
\end{figure}

Fig. \ref{figopt1} shows the optical light curve. Even if the light 
curve is well sampled only in the $R$ band, the general behaviour is 
seen in all the other optical bands. The top panel of 
Fig. \ref{figopt2} is a linear-log plot of the 
$\gamma$-ray/X-ray/optical light curves for an immediate comparison 
with Fig. \ref{xrtbat1}. In the bottom panel of the same figure 
(log-log scale) it is possible to appreciate how the peak in the 
optical band coincides with the deviation in the X-ray band from the 
exponential tail. 
 
 
\setcounter{table}{2} 
\begin{table} 
 \caption{Best-fitting parameters of the optical light curve.} 
 \label{optfit} \begin{tabular}{@{}ccccc} 
\hline 
\hline 
Component & $N^{\rm opt}(F_{j})$ & $\alpha$ & $t_{\rm interval}$ \\ 
  & (mJy) & & (s) \\ 
\hline 
\hline 
$f_{1}$ & ($6.84\pm2.83$) $\times$ 10$^{-6}$ & $-1.56\pm0.70$ & $<460$ \\ 
$f_{2}$ & $5.81\pm3.58$ & $0.61\pm0.09$ & $460<t<1500$ \\ 
$f_{3}$ & ($3.8\pm3.2$) $\times$ 10$^{3}$ & $1.51\pm0.12$ & $1500<t<10^{4}$ \\ 
$f_{4}$ & $0.09\pm0.19$ & $0.41\pm0.17$ & $t>10^{4}$ \\ 
\hline 
\hline 
 \end{tabular} 
\end{table}

\begin{figure} \centering  
\includegraphics[height=8.2cm,width=6.5cm,angle=-90]{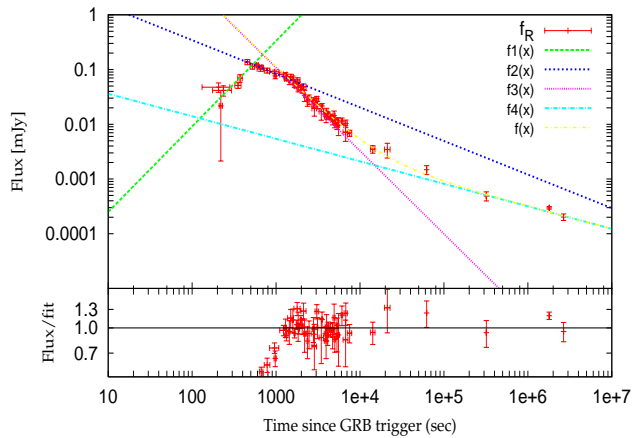} 
\caption{$R$-band optical light curve best-fitting. The fit $f(x)$ is  
the sum of two power laws ($f3$, $f4$) for which the parameters are 
reported in Table \ref{optfit}. The lower panel shows the residuals of 
the fit for the last two components.} 
\label{figopt4} 
\end{figure}

In Fig. \ref{figopt4} we show the simple fit of the $R$-band light 
curve with a series of single power-law segments. Each segment is 
shown together with the sum of the two components at late times. The 
value of the decay index of each power law is reported in Table 
\ref{optfit}, together with the time intervals over which the data has 
been fitted by the correspondent component. The $\chi^{2}$/dof of the 
late-time fit $f(x)$ (after 10$^{3}$~s), consisting of the two 
components $f_{3}$ and $f_{4}$, is 37/41 = 0.91. 
 
 
\subsection{Infrared data and spectral energy distribution} 
 
The fading afterglow of GRB 070419A was detected in the infrared (IR) 
bands thanks to the United Kingdom Infrared Telescope 
(UKIRT). Observations started about 40 min after the event and were 
performed with $JHK$ filters (Rol et al. 2007). The calibrated IR 
magnitudes (with respect to the 2MASS catalogue) are reported in Table 
\ref{obslog}. As for the optical band, IR magnitudes have been 
corrected for the estimated extinction ($A_J$ = 0.025 mag, $A_H$ = 
0.016 mag, $A_K$ = 0.010 mag) and then converted into flux 
densities. Coupling those data with optical and X-ray data, we have 
constructed a spectral energy distribution (SED) 
extrapolating/interpolating the observations at a common time, chosen 
to be $t_{\rm SED}$ = 3000~s = 50 min after the burst. In doing this, 
we excluded the estimate of the flux in the $V$ band, for which the 
calibration of the optical data is uncertain and therefore the 
inferred value for the flux density is not accurate. The extrapolated 
fluxes for all the filters are reported in Table \ref{sedtab}.

 
\begin{figure} \centering  
\includegraphics[height=8.2cm,width=5.2cm,angle=-90]{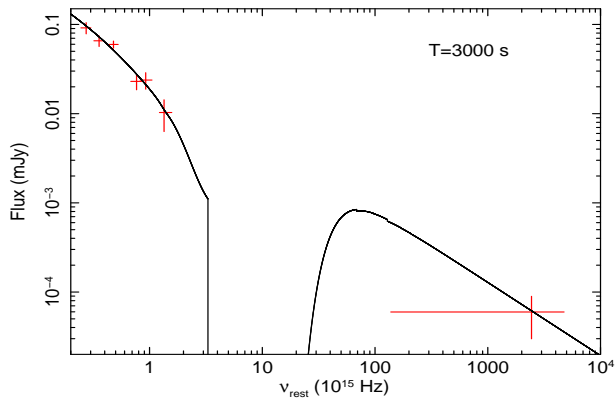} 
\caption{Spectral energy distribution fit at $t=3000$~s after the burst event.} 
\label{figsed1} 
\end{figure}

 
\setcounter{table}{3} 
\begin{table} 
\begin{center} 
\begin{minipage}{126mm} 
 \caption{Observed $F_{\nu}(t = \Delta t)$ and extrapolated $F_{\nu}(t =  
 3000~{\rm s})$ flux. } \label{sedtab} 
 \begin{tabular}{@{}cccc} 
\hline \hline 
Band & $\Delta t$ & $F_{\nu}(\Delta t)$ & $F_{\nu}(3000~{\rm s})$\\ 
 & [s] & [mJy] & [mJy]\\ 
\hline \hline 
$B$ & 2878 & 0.0205 $\pm$ 0.0057 & 0.0103 $\pm$ 0.0040 \\ 
$V$ & 2930 & 0.0392 $\pm$ 0.0039 & 0.0306 $\pm$ 0.0040 \\ 
$R$ & 3063 & 0.0292 $\pm$ 0.0029 & 0.0238 $\pm$ 0.0050 \\ 
$i'$ & 2799 & 0.0194 $\pm$ 0.0058 & 0.0230 $\pm$ 0.0045 \\ 
$J$ & 2930 & 0.0614 $\pm$ 0.0051 & 0.0059 $\pm$ 0.0050 \\ 
$H$ & 2642 & 0.0766 $\pm$ 0.0092 & 0.0065 $\pm$ 0.0090 \\ 
$K$ & 3225 & 0.0838 $\pm$ 0.0014 & 0.0916 $\pm$ 0.0014 \\ 
X-ray & & * & 6.0 $\pm$ 3.0 $\times$ 10$^{-5}$\\ 
\hline \hline 
 \end{tabular} 
\end{minipage} 
*Due to the sparse coverage in the X-ray band after $\sim 10^{3}$~s, 
the extrapolation of the X-ray flux at $t = 3000$~s has been done 
taking into account the two possible fits for the late-time behaviour 
as described in Section 3.2.1. 
\end{center} 
\end{table}

The SED, showed in Fig. \ref{figsed1}, is well fitted by a simple 
optically absorbed power law. The absorption in the X-ray band has 
been fixed to a negligible value because it cannot be determined by 
the fit with a single X-ray point. The assumed extinction profile is 
the Small Magellanic Cloud (SMC) profile (Pei 1992). The best-fitting 
values in the rest frame of the GRB ($z=0.97$) are $\beta_{OX} = 
0.82^{\tiny -0.07}_{\tiny +0.16}$ and $A_V = 0.37 \pm 0.19$ mag.

\section{Discussion} 
 
\subsection{X-ray emission} 
 
From the analysis of the high-energy data of GRB 070419A it seems 
clear that the emission can be simply explained by a central engine 
still active up to at least 250~s, very likely up to 10$^{3}$~s (see 
Fig. \ref{figxrt2}). After that time, the forward-shock emission takes 
over and the X-ray light curve can be explained by a simple power-law 
decay (or a slightly complex two-component model). 
 
\subsection{Optical emission} 
 
Clearly the optical light curve (Fig. \ref{figopt1}) is too complex to 
be explained with the standard forward shock (FS) model. In general, 
optical brightening could be due to the enhancement in the ambient 
density. Since the luminosity above the cooling frequency is 
insensitive to the the ambient density, if the X-ray band is located 
above the cooling frequency, the lack of a corresponding peak in the 
X-ray light curve would be naturally explained.  However, the features 
in the optical light curve are too sharp to be explained by the 
density enhancement model (Nakar \& Granot 2007). Another mechanism 
should be responsible for the production of the observed optical 
features. 
 
A possible mechanism is reverse shock (RS) emission which dominates in 
the optical band at early times.  Although the decay behaviour of 
$\sim t^{-2}$ is well known for the RS emission, the initial decay 
could be as shallow as $t^{-0.5}$ if the typical frequency is above 
the optical band ($\nu_{\rm opt} < \nu_{\rm m,r} < \nu_{\rm c,r}$) at 
the RS crossing time (Sari \& Piran 1999; Kobayashi 2000). Since the 
observed index ($\alpha = 0.61 \pm 0.09$) is consistent with the 
expected value, we test this possibility in detail. 
 
Assuming that the optical peak ($t_{\rm peak} \approx 450$~s) gives 
the fireball deceleration time, we can estimate the initial Lorentz 
factor of the fireball as $\Gamma \approx 350 
n^{-1/8}((1+z)/1.97)^{3/8}(T/450~{\rm s})^{-3/8}(E/(2 \times 10^{51} 
{\rm erg}))^{1/8}$.  If the break in the optical light curve around 
1500~s is due to the passage of the typical frequency of the RS, 
$\nu_{m,r} \propto t^{-54/35}$, through the optical band (Kobayashi 
2000), $\nu_{m,r}$ should be around $3 \times 10^{15}$ Hz at $t = 
t_{\rm peak}$.  The typical frequency of the FS, $\nu_{m,f} = 
\Gamma^{2} \nu_{m,r}$, should be about $4 \times 10^{20}$ Hz at the 
peak time (e.g., Kobayashi \& Zhang 2003).  We expect that the FS 
emission in the optical band should increase as $t^{1/2}$ in the ISM 
(or $t^{-1/4}$ in the wind ambient) until $t \approx 4 \times 
10^{6}$~s when $\nu_{m,f} \propto t^{-3/2}$ passes through the optical 
band. Since the observed optical luminosity already decreases as 
$t^{-0.4}$ or steeper around 10$^{4}$~s, this is not consistent. 
 
If the fireball is magnetized, the $\nu_{m,f}$ could be smaller by a 
factor of $R_{\rm B}^{-1/2}$, where $R_{\rm B}$ is the ratio of the 
microscopic parameters in the two shock regions (we use the same 
notation as Gomboc et al. 2008; $R_{\rm B}$ = $\varepsilon_{\rm 
B,r}/\varepsilon_{\rm B,f}$). The passage time of $\nu_{m,f}$ through 
the optical band scales as $R_{\rm B}^{1/3}$. In order to get a FS 
peak time two orders of magnitude smaller, a very large $R_{\rm B} 
\approx 10^6$ is needed.  With this value of $R_{\rm B}$, the 
luminosity ratio between the RS and FS peaks is about $\Gamma R_{\rm 
B}^{1/2} \approx 4 \times 10^5$. In the observational data, the RS 
component is only a few times brighter than the FS component.  The 
introduction of magnetization cannot fix the problem with the FS 
peak. If we stick with the RS model, the typical frequency of the RS 
emission ($\nu_{m,r}$) should be well below the optical band at the 
shock crossing time. The initial shallow decay ($t^{-0.5}$) in the 
optical might be explained with energy injection to the fireball 
ejecta (refreshed shocks). We cannot rule out this possibility, but 
the energy injection rate should be tuned very carefully in order to 
reproduce the observed power-law decay, and moreover we would need to 
assume a sharp cessation of the injection to get the clear break. 
 
The shallow decay phase observed between 450 and 1500 s, still 
described by a power-law, and the very sharp transitions from one 
power-law to another in the optical light curve are puzzling features 
for any model. One possible explanation of the observed features in 
the optical light curve is the assumption of a finely tuned long-lived 
central engine. This condition is indeed necessary to explain the 
plateau phase observed in the canonical X-ray light curves of many 
GRBs.

\subsection{Supernova/host-galaxy contribution} 
 
As discussed above, the optical light curve continues to decay to 
late times with an unusually shallow gradient. Late-time observations 
consist of LBT detections in the SDSS-$r$ filter, the first ($t \approx 3.2 
\times 10^{5}$~s) and third ($t \approx 2.6 \times 10^{6}$~s) of which 
are consistent with the shallow decay of the afterglow emission. At $t 
\approx 1.8 \times 10^{6}~s$, a small excess above the underlying power law 
is observed (see Fig. \ref{figopt1}). This feature has been attributed 
to a possible supernova component (Dai et al. 2008), but 
interpretation of the late-time data are critically dependent on the 
model of the underlying afterglow. Given the faintness of the emission 
at these late times ($r' \approx 25.7$ mag), which is not atypical for 
GRB host-galaxy magnitudes at this redshift (Wainwright et al. 2007; 
Ovaldsen et al. 2007), future deep optical imaging of the GRB location 
would confirm whether a host-galaxy contribution is present or the 
light has continued to decline below detection limits. 
 
\section{Conclusions} 
 
\begin{itemize} 
 
\item The $\gamma$-ray profile of GRB 070419A consists of a  
single shot (FRED) a few hundred seconds long. The $\gamma$-ray 
fluence ($\sim 5 \times 10^{-7}$ erg cm$^{-2}$) has an average value 
for the {\it Swift} GRB population, but the peak photon flux puts this 
event at the low end of that distribution among {\it Swift} GRBs 
(Sakamoto et al. 2008). 
 
\item The XRT 0.3--10 keV data show the presence of some intrinsic  
(rest-frame) absorption, and there is weak evidence for an evolution 
of both $N_{\rm H}$ and the photon index (mean value of $\Gamma_X \sim 
2.2$) with time. 
 
\item The XRT-BAT fluxes, derived in the 15--150 keV band after  
removing the absorption in XRT curves, shows that the initial steep 
X-ray decay is just the decay of the FRED, modelled with a Norris 
profile (Norris et al. 1996). The early BAT-XRT light curve is due to 
internal shocks or other processes related to prolonged central engine 
activity up to at least 250~s. 
 
\item After 400--500~s, the late-time X-ray power law begins to emerge,  
with index $\alpha_X \approx 1.2 \pm 0.2$ (or $\alpha_{X,2} \approx 
0.7$ in the case of the more complex fit model with two power 
laws). This behaviour seems to be supported by the optical light curve 
that shows a peak roughly at the same time ($t_{\rm peak} \approx 
450$~s). 
 
\item We tried to explain the behaviour of the optical light curve  
in the context of the standard fireball model. However the simple and 
natural explanation (FS plus RS components) does not work. The 
magnetization of the ejecta could affect the FS peak time 
significantly, but the effect is negligible and not in agreement with 
the observations. Another possibility to explain the optical behaviour 
is to argue for the existence of a significant density enhancement in 
the ambient medium. The luminosity below the cooling frequency is 
proportional to $\rho^{1/2}$. If a blast wave hits a density 
enhancement of several tens or hundreds, the bump in the optical at 
$\sim$450~s could be explained. However, the observed peak feature 
might be too sharp for this model. 
 
\item We cannot rule out the possibility of a finely tuned central  
engine for GRB 070419A. However, the optical light curve would not be 
easily reproduced without assuming a particular energy injection rate 
and a sudden cessation of the injection in order to create the 
observed optical features. 
 
\item From our detailed multiwavelength analysis we can conclude that  
GRB 070419A is not explained in the context of the simple standard 
fireball model. Assuming energy injection or a tail component 
following the fireball, or even a more complex emission picture, the 
RS model might explain the observations, but this scenario would 
result in a rather {\it ad hoc} explanation only for this particular 
event. 
 
\end{itemize} 
 
\section*{Acknowledgments} 
 
AM acknowledges funding from the Particle Physics and Astronomy 
Research Council (PPARC). CGM is grateful for financial support from 
the Royal Society and Research Councils UK.  AVF's group is supported 
by US National Science Foundation (NSF) grant AST--0607485, Gary and 
Cynthia Bengier, and the TABASGO Foundation. The Liverpool Telescope 
is operated by Liverpool John Moores University at the Observatorio 
del Roque de los Muchachos of the Instituto de Astrofisica de 
Canarias. The Faulkes Telescopes, now owned by Las Cumbres 
Observatory, are operated with support from the Dill Faulkes 
Educational Trust. KAIT and its ongoing operation were made possible 
by donations from Sun Microsystems, Inc., the Hewlett-Packard Company, 
AutoScope Corporation, Lick Observatory, the NSF, the University of 
California, the Sylvia \& Jim Katzman Foundation, and the TABASGO 
Foundation. We thank R. Chornock for initial discusssions of the KAIT 
data on the optical afterglow of GRB 070419A. This work made use of 
data supplied by the UK {\it Swift} Science Data Centre at the 
University of Leicester.

\label{lastpage} 
  
\end{document}